\documentclass[aps,superscriptaddress,reprint,prl]{revtex4-1}
\usepackage{amsmath}
\usepackage{graphicx, placeins}

\begin{document}

\title{Magnetic coupling at ferromagnetic rare earth / transition-metal interfaces: A comprehensive study}



\author{T. D. C. Higgs}
\affiliation{\mbox{Department of Materials Science and Metallurgy, University of Cambridge, CB3 0FS, Cambridge, United Kingdom}}

\author{S. Bonetti}
\altaffiliation{Current address: Department of Physics, Stockholm University, SE-106 91 Stockholm, Sweden}
\affiliation{\mbox{Department of Physics, Stanford University, Stanford, CA 94305, USA}}

\author{H. Ohldag}
\affiliation{\mbox{SLAC National Accelerator Laboratory, California 94025, USA}}

\author{N. Banerjee}
\affiliation{\mbox{Department of Materials Science and Metallurgy, University of Cambridge, CB3 0FS, Cambridge, United Kingdom}}

\author{X. L. Wang}
\affiliation{State Key Laboratory of Superlattices and Microstructures, Institute of Semiconductors, Chinese Academy of Sciences, Beijing 100083, China}

\author{A. J. Rosenberg}
\affiliation{\mbox{Stanford Institute for Materials and Energy Science, SLAC National Accelerator Laboratory, Menlo Park, California 94025, USA}}
\altaffiliation{Department of Applied Physics, Stanford University, Stanford, California 94305, USA}

\author{Z. Cai}
\affiliation{\mbox{Department of Materials Science and Metallurgy, University of Cambridge, CB3 0FS, Cambridge, United Kingdom}}

\author{J. H. Zhao}
\affiliation{State Key Laboratory of Superlattices and Microstructures, Institute of Semiconductors, Chinese Academy of Sciences, Beijing 100083, China}

\author{K. A. Moler}
\affiliation{\mbox{Stanford Institute for Materials and Energy Science, SLAC National Accelerator Laboratory, Menlo Park, California 94025, USA}}
\altaffiliation{Department of Applied Physics, Stanford University, Stanford, California 94305, USA}

\author{J. W. A. Robinson}
\email[Correspondence to: ]{jjr33@cam.ac.uk}
\affiliation{\mbox{Department of Materials Science and Metallurgy, University of Cambridge, CB3 0FS, Cambridge, United Kingdom}}

\date{\today}

\begin{abstract}
	Thin film magnetic heterostructures with competing interfacial coupling and Zeeman energy provide a fertile ground to study phase transition between different equilibrium states as a function of external magnetic field and temperature.
	A rare-earth (RE) / transition metal (TM) ferromagnetic multilayer is a classic example where the magnetic state is determined by a competition between the Zeeman energy and antiferromagnetic interfacial exchange coupling energy.
	Technologically, such structures offer the possibility to engineer the macroscopic magnetic response by tuning the microscopic interactions between the layers.
	We have performed an exhaustive study of a nickel/gadolinium system by using the element-specific measurement technique x-ray magnetic circular dichroism, and determined the full magnetic state diagrams as a function of temperature and magnetic layer thickness.
	We explain our result based on a modified Stoner-Wohlfarth formalism and provide evidence of a thickness-dependent phase transition to a magnetic fan state which is critical in understanding magnetoresistance effects in RE/TM systems.
	The results provide important insight for spintronics and superconducting spintronics where engineering tunable magnetic inhomogeneity is key for certain applications.
 
\end{abstract}
\maketitle

Modified magnetic interaction at the interfaces between different ferromagnetic materials can be utilised to engineer materials with magnetic properties that are significantly different from the bulk. 
A classical example of such a system is a rare earth / transition-metal (RE/TM) multilayer since in the presence of an external magnetic field the layers favor parallel alignment but the antiferromagnetic coupling between the layers can lead to novel magnetic states and phase transitions between them.
Ferrimagnetic alloys and multilayers (favored antiparallel alignment between layers) are of great interest from both a fundamental and an applied point of view.
Ferrimagnetic alloys were until very recently the only magnetic system that showed ultrafast magnetization switching induced by a femtosecond laser pulse \cite{Mentink2012PRL, Kirilyuk2010RMP, Graves2013NatMat}, and engineered magnetic multilayers are key components of superconducting spin valves \cite{EschrigPhys2011, RobinsonPRL2010, BergeretRMP2005}.

The theoretical groundwork of RE/TM systems was established by Camley and Tilley \cite{CamleyTilleyPRB1988, CamleyPRB1989}; they modelled RE/TM multilayers and found that due to the different Curie temperatures ($T_C$) of the two magnetic components several phases occur as the effects of temperature and applied magnetic field are incorporated.
At low temperatures ($T < T_C(RE)$), the RE aligns with the applied field and the TM aligns antiparallel to the RE and the field (RE-aligned state), while at higher temperatures ($T > T_C(RE)$) the magnetization of the RE decreases and a second-order phase transition to the TM-aligned state occurs.
This work was later confirmed experimentally \cite{BaczewskiJMMM1998, BarthJPCM2008}, however, Camley also predicted that at the interface between bulk Fe and a thin layer of Gd (five atomic layers), a twisted magnetic state could form at some values of applied field \cite{CamleyPRB1987}, although this is still an open question.

Due to the interface-driven nature of RE/TM systems it is difficult to measure the exact magnetic structure of a system with bulk magnetic measurement techniques such as vibrating sample magnetometry and magnetoresistance measurements \cite{PrietoPRL2003, McGrathPRB1996, TakanashiJPSJ1992}.
Additionally, at low temperatures the magnetization of a RE layer in a typical RE/TM bilayer will be far larger than that of the TM layer, making it extremely difficult to deconvolute the two signals.
Therefore, an element specific measurement technique such as x-ray magnetic circular dichroism (XMCD) is ideally suited to measuring such a system.
The energy of the x-rays used can be tuned to the absorption edges of particular elements, meaning only the magnetic response of specific layers is measured at any one time.
Previous studies using XMCD have mainly confirmed the predictions of Camley and Tilley \cite{CamleyTilleyPRB1988, CamleyPRB1989}: Barth \emph{et al.} observed evidence of the transition between the RE- and TM-aligned state as a function of temperature in a Ni/Gd bilayer (7.5 nm and 5 nm thick, respectively) \cite{BarthJPCM2008}, while Koizumi \emph{et al.} observed the same behavior in a Gd/Fe (2 nm/2 nm)$_{50}$ superlattice measured at 20 K and room temperature \cite{KoizumiPRB2000}.

In the present study we present extensive XMCD results of Ni/Gd/Ni films allowing us to fully map the magnetic states as a function of applied field and temperature for a wide range of different Ni and Gd thicknesses (Figs \ref{fig:figure1}(a) and (b)).
Our results demonstrate the full richness of the magnetic phase diagram of RE/TM systems and shed new light on previous work that interpreted results based on an incomplete picture of RE/TM behavior \cite{PrietoPRL2003, RobinsonSciRep2012}.

We measured 15 samples by XMCD, each with a different layer thicknesses, throughout a range of temperatures between 6 K and room temperature
\footnote{These experiments were performed at beam line 13-1 at SLAC National Accelerator Laboratory.
The samples were grown by D.C. magnetron sputtering in ultra-high vacuum conditions (base-pressure better than 10$^{-8}$ mbar).}.
The measured samples displayed one of three behaviors as illustrated in Fig. \ref{fig:figure1}(c) which shows element-specific magnetization versus applied field ($M(H)$) hysteresis loops at high and low temperatures for three different samples that are characteristic of each behavior.
In the first behavior the Ni layers follow the applied magnetic field (positive magnetization in positive field) while the Gd layer remains antiparallel to the Ni and the magnetic field.
This does not change as the temperature is increased.
The hysteresis loops shown are from a sample with total Ni thickness of 5 nm, and Gd thickness of 4.5 nm
\footnote{Note that the multilayers are of the form Ni/Gd/Ni, with equal Ni thicknesses either side of the Gd.  This means that the thicknesses of the individual Ni layers will be half that of the total Ni thickness, and it is the total Ni thickness which is measured and stated in the text and figure captions.}.
In the second behavior, the opposite is observed at lower temperatures: the Ni is antiparallel to the field while the Gd is parallel.
Above a transition temperature, this state is reversed to the same as that in behavior 1.
The thicknesses of the Ni and Gd in this sample are 6 nm and 8 nm, respectively. 
These two behaviors can be understood in terms of the variation in the influence of the interfacial layers in the multilayer compared to that of the bulk.
In the thinnest samples, the interfacial coupling will dominate and if the ratio between the Ni and Gd thicknesses favors the Ni, the Ni will control the behavior of the Gd despite the higher magnetic moment of the Gd at lower temperatures.
For slightly thicker Gd layers, although the interfacial coupling still dominates the behavior of the system, the Gd will dominate the Ni while its magnetic moment is not suppressed by the increasing temperature.

The third behavior is more complex, with the Ni $M(H)$ loops at low temperature displaying a distinctive ``N''-shape, indicative of significant magnetic inhomogeneity in Ni during magnetization reversal. 
Meanwhile, the Gd layer undergoes a gradual transition over a temperature range of tens of kelvins from aligned parallel to the Ni to antiparallel.
The sample displaying this behavior in the figure had Ni and Gd thicknesses of 8 nm and 6 nm, respectively, and is the same sample as that for which the XMCD spectra are shown in Fig. \ref{fig:figure1}(b).
We do not believe this behavior in the Ni has been observed before; the magnetization crosses zero three times on each branch of the hysteresis loop. 

In order to understand the micromagnetic structure that corresponds to these hysteresis loops, we have modelled the TM/RE/TM multilayer.
The model uses the Stoner-Wohlfarth framework by Mauri \emph{et al.} \cite{MauriJAP1987}, and that of Camley and Tilley, wherein an iterative process is used to find the ground state of a stack of Ni and Gd atomic layers.
However, our model includes a term describing the magneto-crystalline anisotropy energy since some previous work had not been able to use Camley and Tilley's model to explain the behavior of an RE/TM multilayered system \cite{HahnPRB1995}, citing the possibility that only including the Zeeman and exchange energy terms leads to limitations in explanatory power. 
We solve the following Hamiltonian to find the angle of one spin at which the energy is a minimum:

\begin{align}
	E_i = &- g \mu_B \mu_0 M_i H \cos(\phi_i) + \frac{K_i}{n} \sin^2(\phi_i - \theta) \nonumber \\
				&- J_{i+1} S_{i+1} S_i \cos(\phi_i - \phi_{i+1}) \nonumber \\
				&- J_{i-1} S_{i-1} S_i \cos(\phi_i - \phi_{i-1}),
	\label{eqn:energy_tot}
\end{align}

where $g$ is the Land\'e g-factor, $\mu_B$ is the Bohr magneton, $\mu_0$ is the permeability of free space, $M$ is the average magnetization of the layer (in our case $M$ is equal to the value of the magnetic moment of the atom for a perfectly ordered material), $\phi$ is the angle of a magnetic moment with respect to the applied magnetic field direction, $K$ is the magnetocrystalline anisotropy energy, $\theta$ is the direction of the magnetic easy axis, $n$ is the number density of the material of the layer, $J$ is the exchange constant of the layer, and $S$ is the magnitude of the magnetic moment of an atom in units of the Bohr magneton.
We then repeat the process for each spin in the Ni/Gd/Ni stack until successive iterations over the whole stack differ by no more than a factor of $1/1000$.
A detailed description of the model can be found in the Supplemental Material.
Figure \ref{fig:arrows} shows a comparison between the results of the model and the temperature dependence of the third type of behavior, which match very closely despite the simplicity of the model.
Although the model can reproduce all three of the behaviors discussed above, here we focus only on the thrid type.
It should be noted that one of the few assumptions made for the model is that of strict antiferromagnetic coupling (negative exchange energy) between the Ni and Gd layers.
The success of this assumption confirms that no more complicated effects are influencing the behavior of the interface in order to understand the complex $M(H)$ loops.
We can see that at low temperatures, the Ni magnetization increases with increasing field, but abruptly reverses as the field crosses zero, before saturating.
This behavior is reproduced by the model, as is the thermal evolution of the hysteresis loops.
As the temperature increases, the ``N''-shape opens up until at $\sim$ 30 K the Ni loop looks more conventional, except with inflection points on the loop at the same field value as the inflection points at low temperatures.

In the Gd measurements, we can now see the details of the transition from aligning parallel to aligning antiparallel with the field.
The saturation magnetization, $M_s$, at negative field increases while the opposite is true for the $M_s$ at positive fields.
Interestingly, for most of the temperature range shown the saturation magnetization is not the highest value of magnetization within the loop; at saturation there must be significant Gd moments canted away from the applied field direction.
This feature, combined with the unusual shape of the Ni loops leads us to conclude that some form of magnetic inhomogeneity exists in the Ni/Gd/Ni structure.

The model we use allows us to extract the micromagnetic structure from the system that produced the hysteresis loops since the modelled system is a one-dimensional stack of magnetic moments.
We assume that the moments in each layer are parallel, and so can apply a normalisation factor to each layer's magnetization.
This helps to reduce the computational complexity of the model, and since the measured samples are polycrystalline, film orientation effects will be negligible.
The model does not take into account the crystal structure since in a polycrystalline film orientation effects will be negligible.
The sum of the projections of the magnetic moments of each of the elements onto the direction of the applied field gives the total magnetization at each field value, as shown in the hysteresis loops.
But the original angles of the each magnetic moment can also be plotted individually to give vector diagrams that show the inhomogeneity in the system at particular field values.

These vector diagrams are shown in Fig. \ref{fig:arrows}(b).
They show the angles of the magnetization for each atomic layer calculated by the model at a particular field value and temperature; the field values are indicated on the corresponding hysteresis loops.
From the vector diagrams we can see that the reason for the shape of the Ni hysteresis loops at low temperature is the effect the Ni/Gd interface has on the Ni layers.
At saturation fields the Ni magnetization is not at its maximum value because the moments at the interface have been canted away from the field direction due to the competition between the inter-, intra-layer exchange energiesand the Zeeman energy.
The sharp decrease in magnetization in the Ni at small positive fields is due to the Gd exerting its influence on the Ni when the Gd reverses.
The reversal of the dominant Gd ``drags'' the Ni round to remain antiparallel to the field before the increasing Zeeman energy of the Ni at increasing fields forces the degree of winding near the interface to be suppressed.
By examining the model in this way, a unique understanding of the system and its magnetic microstructure can be gained.

As well as yielding information about the magnetic microstucture of the system, the model can be used to extract other magnetic properties.
For example, one fitting parameter in the model is the $T_C$ of the Gd.
Due to the large difference in the Curie temperatures between RE and TM ferromagnets, and the temperatures at which the experiment was performed, the value of the Ni saturation magnetization as a function of temperature in the model is presumed to be constant.
However, since the $T_C$ of bulk Gd is very close to room temperature, the decrease in magnetization as a function of temperature must be taken into account in the model, and the effective $T_C$ of Gd will be further suppressed due to the geometry of the Gd layer.
The value of $T_C$ can be extracted from the model; Figure \ref{fig:MvsT}(a) shows the agreement between the experiment and the model when the $T_C$ of Gd in the model is set to 35 K.
The suppression of the Curie temperatures of ferromagnets as a function of film thickness is well documented \cite{JiangJAP1996, ZhangPRL2001, JiangPRL1995}.
The $\Delta M_s$ values were calculated by averaging the maximum and minimum magnetization values within each loop and taking the difference between them.
It should be noted that the magnetization values have been normalised to between zero and one, and that therefore the rise in the Ni magnetization with temperature is not comparable to the decrease of Gd magnetization.
In fact, the rise in the Ni originates due to the method of measurement; XMCD only detects the projection of a sample's magnetization onto the $\bf k$-vector of the x-ray, so as more Ni moments are able to align with the applied field ,which is parallel to the x-ray's $\bf k$-vector, the measured magnetization will increase.

Figure \ref{fig:MvsT}(b) shows a magnetic state diagram for a particular sample, that can be extracted from measured hysteresis loops due to their element-specific nature.
The Ni and Gd hysteresis loops are first normalized, so the maximum of the modulus of the difference is 2 ($\Delta M_s = 2$); this corresponds to the Ni and Gd having oppositely saturated magnetizations, while $\Delta M_s$ = 0 corresponds to parallel Ni and Gd.
We know that when the Ni and Gd layers are antiparallel there is no inhomogeneity at the interfaces, but the inhomogeneity will nucleate at the interfaces as the angle between the layers changes.
Magnetic state diagrams such as these will be invaluable in the design of devices in a variety of fields that require engineered magnetic inhomogeneity.

We must point out that in all the Figures showing the output of the model, the magnetic field values that can be compared with those from the experiment, coercive field values for instance, are larger than their counterparts by an approximately constant factor of 4.
This difference arises due to the fact that simple models do not include nucleation points present in physical samples that facilitate magnetization reversal at lower magnetic field values \cite{HartmannPRB1987, BrownRMP1945, BalasSciRep2014, AharoniRMP1962}.
This discrepancy should not affect the results of the model in any way other than a scaling of the equivalent fields. 

The measured samples were grown in four different growth runs, but we are confident that the observed effects are due to thickness dependence, and not stochastic variations in the growth conditions.
Firstly, because samples from different growth runs display the same types of behaviors.
Secondly, imaging the magnetic landscape on a micron-scale for three samples using a scanning SQUID microscope \cite{HuberRSI2008}, a sensitive probe of magnetic flux, shows consistent resolution-limited domain structure and uniform magnetization strength for a variety of location in each sample (see Supplemental Material for more information).

In summary, we have performed a comprehensive experimental and theoretical study of the RE/TM system, using both a more sophisticated model than has previously been utilised and an extremely powerful measurement technique (XMCD).
The use of the model allows a detailed view of the micromagnetic structure that is responsible for any of the three behaviors that were observed.
The strict antiferromagnetic exchange coupling between RE and TM ferromagnets and its competition with the intralayer exchange energy and the Zeeman energy has been shown to lead to significant inhomogeneity in the system.
The relative strengths of these energies, and thus the behavior of the system can be tuned by varying the thickness of the individual layers within the multilayer, and our characterisation of the necessary thicknesses should enable the design of a new generation of devices that can utilise the magnetic inhomogeneity present at RE/TM interfaces including spintronics and superconducting spintronics.
Our work will help to reinterpret the results of previous studies, and has consolidated our knowledge of another otherwise simple system that can be engineered to yield not only fundamentally interesting behaviors, but behaviors that have the potential to influence future technologies based on spintronics and superconducting spintronics \cite{Linder2015NatPhys}.

\FloatBarrier

\begin{widetext}
	\begin{figure}
		\includegraphics[width=\textwidth]{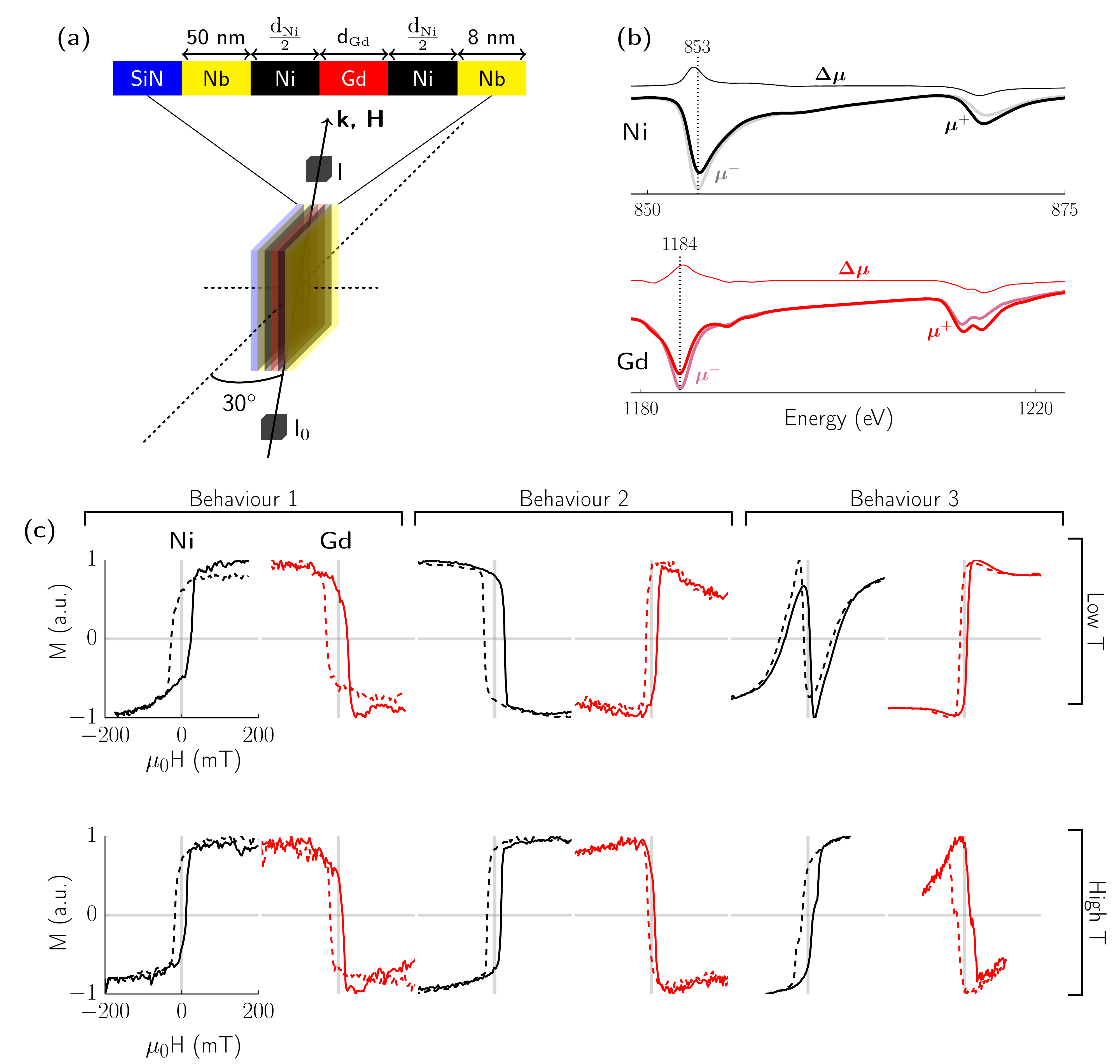}
		\caption{(a) Diagram of experimental setup showing the structure of the samples, the direction of the incident x-rays, and the position of the measuring photodiodes ($I_0$ and $I$). (b) Examples of XMCD spectra taken at $\sim$6 K for Ni (i) and Gd (ii). The XMCD was measured at the Ni $L_3$ and Gd $M_5$ absorption edges, respectively, marked by dotted lines. The two signals $\mu^+$ and $\mu^-$ are from positive and negative saturation, while $\Delta \mu$ is the difference between the two, vertically offset for clarity. (c) Hysteresis loops measured between 5 and 10 K (low T), and 45 and 65 K (high T), that are representative of the three observed behaviors. The (total) Ni and Gd thicknesses of each of the samples are (from left to right): 5 nm Ni and 4.5 nm Gd, 6 nm Ni and 8 nm Gd, and 8nm Ni and 6 nm Gd. The last sample is the same as that from which the XMCD spectra are shown in (b).}
		\label{fig:figure1}
	\end{figure}

	\begin{figure}
		\includegraphics[scale=0.4]{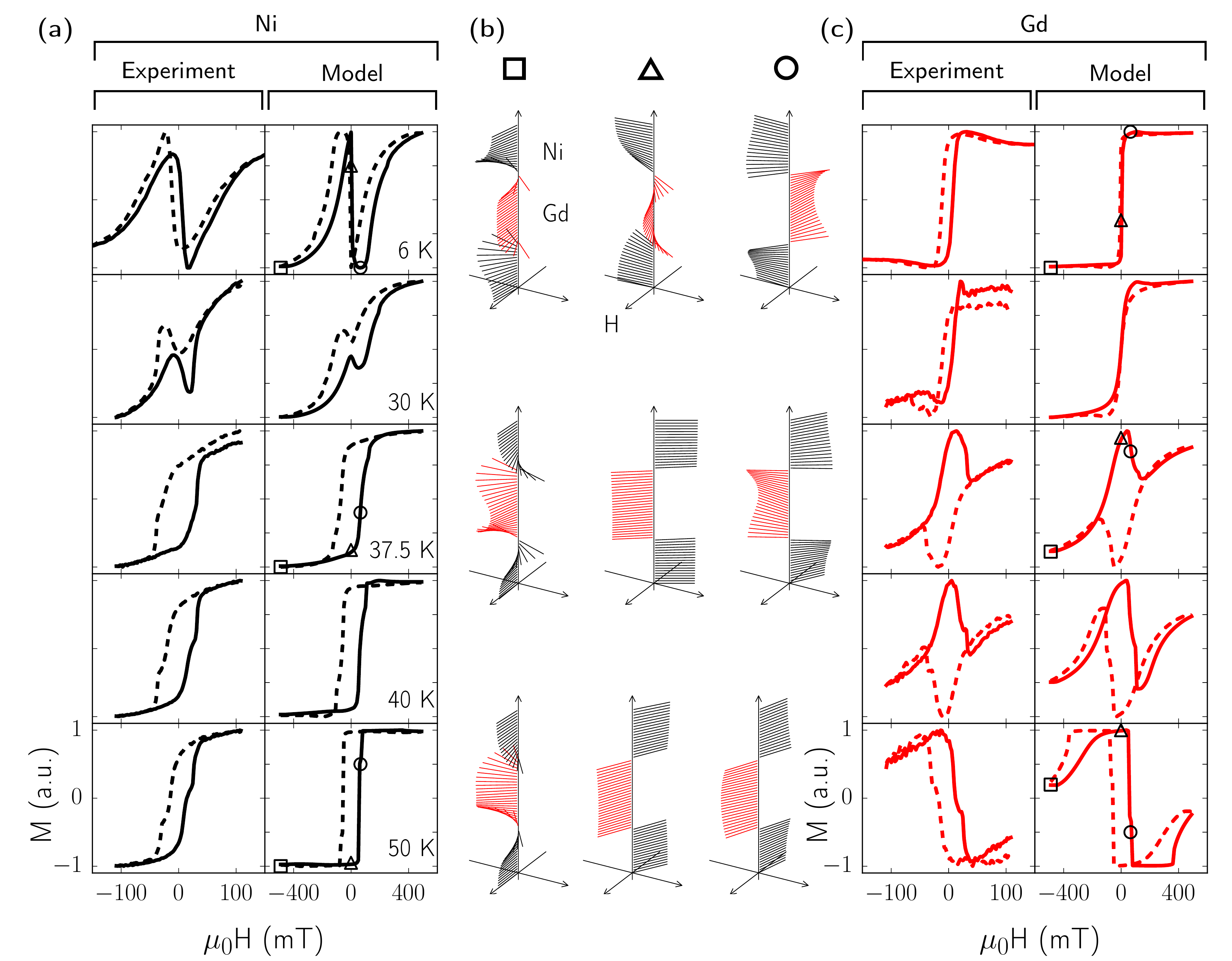}
		\caption{Comparison between the hysteresis loops measured by XMCD and those produced by the model, as well as detailed view of the micromagnetic structure of the sample at certain temperatures and field values extracted from the model. (a, c) Hysteresis loops, experimental and theoretical for Ni and Gd, respectively.  The magnetization has been normalised by the highest and lowest values of magnetization within the loop to ease comparison.  Shaped markers on some theoretical loops show the temperature and field values of the vector diagrams in (b).  (b) Vector diagrams showing the direction of magnetization of each atomic layer in the model system used to produce the hysteresis loops.  The magnitude of the magnetization is not shown.  Black arrows represent Ni layers, red Gd, as labelled in the top left. The experimental hysteresis loops are from the same sample as that shown in Fig. \ref{fig:figure1}(b) and (c)(behavior 3), and has (total) Ni and Gd thicknesses of 8 nm and 6 nm, respectively.}
		\label{fig:arrows}
	\end{figure}
\end{widetext}

\begin{figure}
  \centering
  \includegraphics[scale=0.5]{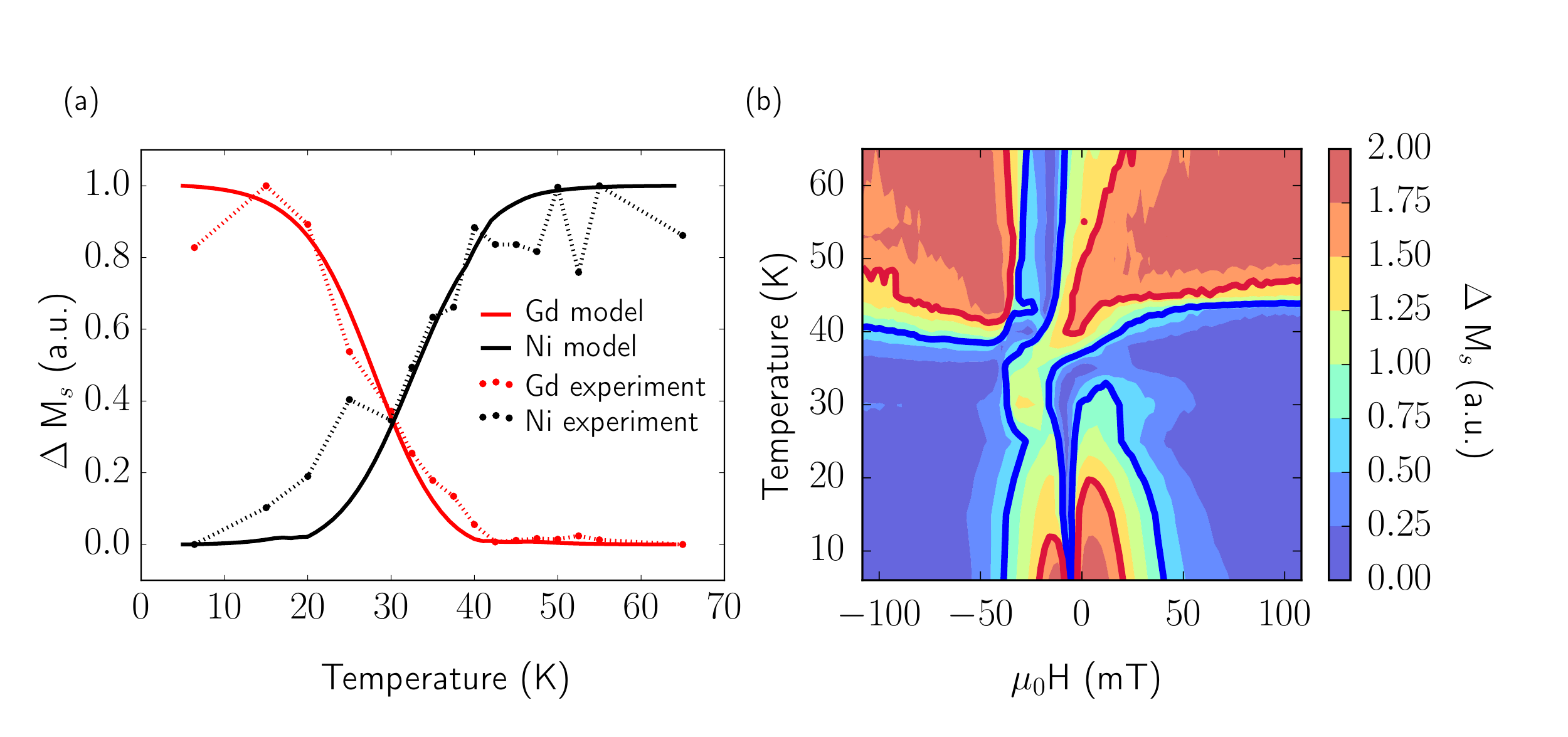}
	\caption{(a) Difference between the average maximum and minimum magnetizations in experimental and theoretical hysteresis loops. The connected dots represent experimental results, while solid lines represent the results of the model.  The plot is generated by taking half of the difference between the average maximum and minimum values of magnetization in a hysteresis loop at a particular temperature.  The whole plot has been normalised to its own maximum and minimum for ease of comparison.
	(b) Phase diagram showing the degree of inhomogeneity present in an experimental multilayer as a function of applied field and temperature. The phase diagram is constructed from experimental hysteresis loops by taking the difference between the normalised magnetization of Ni and Gd at each field value in a loop.  This is then repeated for every temperature.}
  \label{fig:MvsT}
\end{figure}

\begin{acknowledgments}
	T.D.C.H. and J.W.A.R. acknowledge funding from the EPSRC [EP/I038047/1] and the Leverhulme Trust [IN-2013-033].
	S.B. acknowledges support from the Knut and Alice Wallenberg Foundation. 
	X.L.W. and J.H.Z. acknowledge support from the MOST of China [2015CB921500].
	Research at SLAC and Stanford was supported through the Stanford Institute for Materials and Energy Sciences (SIMES) which like the SSRL user facility and the scanning SQUID microscopy is funded by the Office of Basic Energy Sciences of the U.S. Department of Energy.
\end{acknowledgments}


%

\end{document}